\documentclass[12pt]{article}
\usepackage{graphics}
\usepackage{epsfig}
\usepackage{amssymb,amsmath}
\usepackage{array}

 \makeatletter
\def\alt{\mathrel{\mathpalette\gl@align<}}
\def\agt{\mathrel{\mathpalette\gl@align>}}
\def\gl@align#1#2{\lower.6ex\vbox{\baselineskip\z@skip\lineskip\z@
\ialign{$\m@th#1\hfil##\hfil$\crcr#2\crcr\sim\crcr}}} \makeatother

\begin{document}

\vspace*{1.0cm}
\begin{center}
\baselineskip 20pt
{\Large\bf New Fermions at the  LHC and Mass of the Higgs Boson}
\vspace{1cm}

{\large
Ilia Gogoladze\footnote{E-mail:
ilia@physics.udel.edu\\ \hspace*{0.5cm} On  leave of absence from:
Andronikashvili Institute of Physics, GAS, 380077 Tbilisi, Georgia.},
Bin He \footnote{E-mail:
hebin@udel.edu}
and
Qaisar Shafi}

\vspace{1.0cm}

{

\it Bartol Research Institute, Department of Physics and Astronomy, \\
University of Delaware, Newark, DE 19716, USA }

\vspace{.5cm}

\vspace{1.5cm} {\bf Abstract}
\end{center}

Unification at $M_{\rm GUT}\sim 3\times 10^{16}$  GeV of the three Standard Model (SM) gauge couplings can be achieved by postulating the existence of a pair of vectorlike fermions carrying SM charges and masses of order 300 GeV -- 1 TeV. The presence of these fermions significantly modifies the vacuum stability and perturbativity bounds on the mass of the SM Higgs boson. The new vacuum stability bound in this extended SM is estimated to be 117 GeV, to be compared with the SM prediction of about 128 GeV. An upper bound of 190 GeV is obtained based on perturbativity arguments. The impact on these predictions of type I seesaw physics is also discussed. The discovery of a relatively `light' Higgs boson with mass $\sim$ 117 GeV could signal the presence of new vectorlike fermions within reach of the LHC.

\thispagestyle{empty}
\newpage


\baselineskip 18pt

\section{Introduction}

The discovery of the Standard Model (SM) Higgs boson is arguably the
single most important mission for the LHC. Under a
somewhat radical assumption that the next energy frontier
lies at the reduced Planck scale ($M_P\simeq 2.4\times 10^{18}$ GeV), it has been found that the SM Higgs
boson mass lies in the range 128 GeV $\lesssim m_H \lesssim $ 175 GeV
\cite{stability1}. Here the lower bound of 128 GeV on $m_H$ derives
from arguments based on the stability of the SM vacuum. More
precisely, that the Higgs quartic coupling does not become negative at
any scale between $M_Z$ and $M_P$. The upper bound of 175 GeV or so on
$m_H$ stems from the requirement that the Higgs quartic coupling
remains perturbative and does not exceed ${4 \pi}$, say, during
its evolution between $M_Z$ and $M_P$.
Thus, it would appear that discovery of
 a relatively `light' Higgs boson (with mass well below 128 GeV)
 may signal the presence of physics beyond the SM.

 Supersymmetry is by far the most compelling extension of the SM and its minimal realization (MSSM) predicts a relatively `light' SM--like Higgs boson with mass $\lesssim$ 130 GeV. However, in the light of LHC, plausible alternatives to supersymmetry deserve careful investigation. For instance, it was shown in \cite{Gogoladze:2008gf} that the new physics  between $M_Z$ and $M_P$ associated with type II seesaw \cite{seesawII} around TeV scale or higher can yield a `light' Higgs boson with mass $\gtrsim$ 114.4 GeV, the LEP II bound.  A `light' Higgs boson is also realized in scenarios of gauge--Higgs unification with a compactification scale below $M_P$ \cite{Gogoladze:2007qm}.

 In this paper we revisit another extension of the SM, proposed several years ago, in which new TeV scale vectorlike fermions are introduced in order to implement unification at some scale $M_{\rm GUT}$ of the three SM gauge couplings \cite{Amaldi:1991zx}. The new vectorlike fermions carry SM gauge quantum numbers and their presence therefore modifies the SM Higgs mass bounds based on vacuum stability and perturbativity arguments. In particular, by including only a pair of vectorlike fermions for which case $M_{\rm GUT}\simeq 3\times 10^{16}$ GeV, the vacuum stability bound can be lowered from its conventional value of around 128 GeV to a significantly lower value of about 117 GeV.  To keep the discussion as realistic as possible, we also study the possible impact neutrino oscillation physics could have on the Higgs mass predictions. We employ type I seesaw for these considerations \cite{seesawI}. Note that a more complicated scenario containing several new particles (including scalars) can yield $M_{\rm GUT}\simeq M_P$, with a vacuum stability bound as low as 114 GeV.

 This letter is organized as follows. In Section 2, we investigate the impact of new vectorlike fermions on the vacuum stability and perturbativity Higgs mass bounds. In Section 3,
 we discuss how neutrino physics utilizing type I seesaw can affect the vacuum stability and perturbativity bounds. Our conclusion is outlined in section 4.

\section{New Fermions and the Higgs Boson Mass}

We introduce the following vectorlike fermions to achieve gauge coupling unification:
\begin{eqnarray}
Q\left(3, 2, \frac{1}{6}\right) +\overline{Q}\left(\bar3, 2, -\frac{1}{6}\right)+ D\left(3, 1, \frac{1}{3}\right)+ \overline{D}\left(\bar 3, 1, -\frac{1}{3}\right),
\label{nf1}
\end{eqnarray}
where the brackets contain the $SU(3)_c \times SU(2)_L \times U(1)_Y$ quantum numbers of the new particles.
The SM Lagrangian is supplemented by additional terms, and the relevant ones are given by
\begin{eqnarray}
 {\cal L}_{new}= -\kappa_{1} \bar {Q} \bar {D} \Phi^c - \kappa_{2}{Q}  D \Phi -y^i_1Q d_i^c\Phi-y_2^i q_i D \Phi-y^i_3 Q u_i^c\Phi^c - M_{F}(\bar{Q}Q+\bar{D} D) + h.c.
\label{nflag}
\end{eqnarray}
where $\Phi$ denotes the SM higgs doublet,  $\Phi^c\equiv i \sigma_2 \Phi^*$ its charge conjugate, and we employ the standard notation  $q_i,\, u_i^c, \, d^c_i$  for the SM quarks, with $i=1,\,2,\,3$. The parameters $y_{1,2,3}^i$ and $\kappa_{1,2}$ are dimensionless couplings.  We assume, for simplicity, that the new fermions have a common vectorlike mass $M_{F}$.
As pointed out in \cite{Choudhury:2001hs}, most of the $y^i_{1,2,3}$ couplings have to be very small due to constraints from the precision electroweak  data.  To accommodate this, we will assume that the couplings $y^i_{1,2,3}$ are sufficiently small so that they do not give a significant contribution in the RGE analysis. However, the $y^i~'s$ allow the new fermions to decay into the SM particles, without creating any cosmological problems.

There are constraints on the $\kappa_{1,2}$ couplings and the masses of the new matter fields. The most important ones arise
from the $S$ and $T$ parameters which severely limit the number of additional {\it}{chiral}\/ generations.
Consistent with these constraints, one should therefore add new matter which is predominantly vectorlike.
In the limit where the vectorlike mass $M_{F}$ is much heavier than the chiral mass term
(arising from Yukawa coupling to the Higgs doublets), the contribution to the $T$ parameter
from a single chiral fermion is given by~\cite{Lavoura:1992np}
\begin{equation}
\delta T\approx\frac{ N (\kappa_i v)^2}{10 \pi \sin^2\theta_W m^2_W}\left[ \left(
\frac{\kappa_i v}{M_V}\right)^2  +O\left( \frac{\kappa_i
v}{M_V}\right)^4\right], \label{tpar}
\end{equation}
where $\kappa_i$, $i=1,\,2$, are the  Yukawa couplings in Eq. (\ref{nflag}), $v=246.2$ GeV is the
vacuum expectation value (VEV) of the Higgs field, and $N$ counts the number of additional SU(2) doublet pairs, which in our case is $3$. From the precision electroweak data $T\leq 0.06(0.14)$ at 95\% CL for
$m_H=117$ GeV ($300$ GeV) ~\cite{PDG}. We
will take $\delta T<0.1$ as a conservative bound for our analysis.
We see from Eq.~(\ref{tpar}) that with $M_{F}\sim$ 500 GeV, the
Yukawa couplings $\kappa_{i}$ can be $O(1)$.

For the SM gauge coupling we employ the two  renormalization group equation (RGE) \cite{RGE} :
\begin{eqnarray}
 \frac{d g_i}{d \ln \mu} =
 \frac{b_i}{16 \pi^2} g_i^3 +\frac{g_i^3}{(16\pi^2)^2}
  \left( \sum_{j=1}^3 B_{ij}g_j^2 - C^t_i y_t^2   \right),
\label{gauge}
\end{eqnarray}
 where $g_i$ ($i=1,2,3$) are the SM  gauge couplings and $y_t$ is the top Yukawa coupling,
\begin{eqnarray}
b^{SM}_i = \left(\frac{41}{10},-\frac{19}{6},-7\right),~~~~
 { B^{SM}_{ij}} =
 \left(
  \begin{array}{ccc}
  \frac{199}{50}& \frac{27}{10}&\frac{44}{5}\\
 \frac{9}{10} & \frac{35}{6}&12 \\
 \frac{11}{10}&\frac{9}{2}&-26
  \end{array}
 \right),  ~~~~
C^t_i=\left( \frac{17}{10}, \frac{3}{2}, 2 \right).
\label{beta} \end{eqnarray}
For a renormalization scale $\mu > M_{F}$, the beta function for gauge couplings receives an additional  contribution from the vectorlike fermions,
\begin{eqnarray}
b^{\prime}_i = \left(\frac{2}{5}, 2, 2\right),~~~~
 { B^{\prime}_{ij}} =
 \left(
  \begin{array}{ccc}
  \frac{3}{50}& \frac{3}{10}&\frac{8}{5}\\
 \frac{1}{10} & \frac{49}{2}&8 \\
 \frac{1}{5}&3 & \frac{114}{3}
  \end{array}
 \right),  ~~~~
C^{\kappa_1}_i=C^{\kappa_2}_i=\left( \frac{1}{2}, \frac{3}{2}, 2 \right),
\label{beta2} \end{eqnarray}
where $C^{\kappa_1}_i$ and $C^{\kappa_2}_i$ stand for the contribution which is proportional to the $\kappa_i$ coupling in the two loop RGE for  gauge couplings.

For the top Yukawa coupling, we have \cite{RGE}
\begin{eqnarray} \label{ty}
 \frac{d y_t}{d \ln \mu}
 = y_t  \left(
 \frac{1}{16 \pi^2} \beta_t^{(1)} + \frac{1}{(16 \pi^2)^2} \beta_t^{(2)}
 \right).
\end{eqnarray}
Here the one-loop contribution is
\begin{eqnarray}
 \beta_t^{(1)} =  \frac{9}{2} y_t^2 -
  \left(
    \frac{17}{20} g_1^2 + \frac{9}{4} g_2^2 + 8 g_3^2
  \right) ,
\label{topYukawa-1}
\end{eqnarray}
while the two-loop contribution is given by
\begin{eqnarray}
\beta_t^{(2)} &=&
 -12 y_t^4 +   \left(
    \frac{393}{80} g_1^2 + \frac{225}{16} g_2^2  + 36 g_3^2
   \right)  y_t^2 + \frac{1187}{600} g_1^4  - \frac{9}{20} g_1^2 g_2^2 \nonumber \\
 &&
  +\frac{19}{15} g_1^2 g_3^2
  - \frac{23}{4}  g_2^4  + 9  g_2^2 g_3^2  - 108 g_3^4 + \frac{3}{2} \lambda^2 - 6 \lambda y_t^2 .
\label{topYukawa-2}
\end{eqnarray}
In solving Eq.~(\ref{ty}),
 the initial top Yukawa coupling at $\mu=M_t$
 is determined from the relation
 between the pole mass and the running Yukawa coupling
 \cite{Pole-MSbar,Pole-MSbar2},
\begin{eqnarray}
  M_t \simeq m_t(M_t)
 \left( 1 + \frac{4}{3} \frac{\alpha_3(M_t)}{\pi}
          + 11  \left( \frac{\alpha_3(M_t)}{\pi} \right)^2
          - \left( \frac{m_t(M_t)}{2 \pi v}  \right)^2
 \right),
\end{eqnarray}
 with $ y_t(M_t) = \sqrt{2} m_t(M_t)/v$ and $\alpha_3\equiv g_3^2/4\pi$.
Here, the second and third terms in parentheses correspond to
 one- and two-loop QCD corrections, respectively,
 while the fourth term comes from the electroweak corrections at one-loop level.
The numerical values of the third and fourth terms
 are comparable (their signs are opposite).
The electroweak corrections at two-loop level and
 the three-loop QCD corrections
 are both comparable and of sufficiently small magnitude \cite{Pole-MSbar2}
 to be safely ignored.

For a renormalization scale $\mu > M_{F}$, according to the Eq. (\ref{nflag}), the beta function for the top Yukawa coupling receives an additional  contribution at one loop level as follows:
\begin{eqnarray}
 \delta \beta_t^{(1)} =  3(\kappa_1^2+\kappa_2^2),
\label{top2}
\end{eqnarray}
and the additional two loop contributions are
\begin{eqnarray}
 \delta \beta_t^{(2)} =  \left(\frac{5}{8}g_1^2+\frac{45}{8}g^2_2+20g^2_3 \right) (\kappa_1^2+\kappa^2_2)-\frac{27}{4}(\kappa_1^4+\kappa^4_2)
-\frac{27}{4}y^2_t(\kappa_1^2+\kappa^2_2).
\label{top3}
\end{eqnarray}

The one and two loop RGEs  for  the Yukawa couplings $\kappa_1$ and $\kappa_2$ are given by
\begin{eqnarray} \label{kappa5}
 \frac{d \kappa_1}{d \ln \mu}
 = \kappa_1  \left(
 \frac{1}{16 \pi^2} \beta_{\kappa_1 }^{(1)} + \frac{1}{(16 \pi^2)^2} \beta_{\kappa_1}^{(2)}
 \right).
\end{eqnarray}
Here the one loop contribution is
\begin{eqnarray} \label{ty-1}
 \beta_{\kappa_1}^{(1)} = -\frac{1}{4}g_1^2-\frac{9}{4}g^2_2-8g^3+\frac{9}{2}\kappa_1^2+3\kappa^2_2+3y_t^2,
 \end{eqnarray}
while the two-loop contribution is given by
%
\begin{eqnarray}
\beta_{\kappa_1}^{(2)} &=&- \frac{127}{600}g_1^4-\frac{23}{4}g_2^4-108g_3^4
-\frac{27}{20}g_1^2g_2^2+\frac{31}{15}g_1^2g_3^2+9g_2^2g_3^2-
6\lambda \kappa_1^2\nonumber \\
&&+\left (\frac{85}{40}g_1^2+\frac{45}{8}g_2^2+20g_3^2\right )
y _t^2
+\left (\frac{237}{80}g_1^2+\frac{225} {16}g_2^2+36g_3^2
\right ){\kappa_1}^2+\frac{3}{2}\lambda^2 \nonumber \\
&&
+\left (\frac{5} {8}g_1^2+\frac{45}{8}g_2^2+20g_3^2\right )\kappa_2^2
-12\kappa_1^4-\frac{27}{4}(y_t^4+\kappa_2^4+y^2_t \kappa^2_1+\kappa_1^2\kappa^2_2).
\label{dybsmdt}
\end{eqnarray}
The RGE for the Yukawa coupling $\kappa_2$ is obtained by making the replacement  $\kappa_1\leftrightarrow\kappa_2$ in Eqs.  (\ref{kappa5})-(\ref{dybsmdt}). This follows from the various quantum numbers listed in Eq. (\ref{nf1}). As previously mentioned, we are neglecting mixing terms involving the new vectorlike particles and the SM ones.

The RGE  for the Higgs boson quartic coupling is given by \cite{RGE}
\begin{eqnarray}
\frac{d \lambda}{d \ln \mu}
 =   \frac{1}{16 \pi^2} \beta_\lambda^{(1)}
   + \frac{1}{(16 \pi^2)^2}  \beta_\lambda^{(2)},
\label{self5}
\end{eqnarray}
with
\begin{eqnarray}
 \beta_\lambda^{(1)} &=& 12 \lambda^2 -
 \left(  \frac{9}{5} g_1^2+9 g_2^2  \right) \lambda
 + \frac{9}{4}  \left(
 \frac{3}{25} g_1^4 + \frac{2}{5} g_1^2 g_2^2 +g_2^4
 \right) + 12 y_t^2 \lambda  - 12 y_t^4 ,
\label{self-1}
\end{eqnarray}
and
\begin{eqnarray}
  \beta_\lambda^{(2)} &=&
 -78 \lambda^3  + 18 \left( \frac{3}{5} g_1^2 + 3 g_2^2 \right) \lambda^2
 - \left( \frac{73}{8} g_2^4  - \frac{117}{20} g_1^2 g_2^2
 - \frac{1887}{200} g_1^4  \right) \lambda - 3 \lambda y_t^4
 \nonumber \\
 &&+ \frac{305}{8} g_2^6 - \frac{867}{120} g_1^2 g_2^4
 - \frac{1677}{200} g_1^4 g_2^2 - \frac{3411}{1000} g_1^6
 - 64 g_3^2 y_t^4 - \frac{16}{5} g_1^2 y_t^4
 - \frac{9}{2} g_2^4 y_t^2
 \nonumber \\
 && + 10 \lambda \left(
  \frac{17}{20} g_1^2 + \frac{9}{4} g_2^2 + 8 g_3^2 \right) y_t^2
 -\frac{3}{5} g_1^2 \left(\frac{57}{10} g_1^2 - 21 g_2^2 \right)
  y_t^2  - 72 \lambda^2 y_t^2  + 60 y_t^6.
\label{self-2}
\end{eqnarray}
We calculate the Higgs boson pole mass $m_H$ from the running Higgs quartic coupling
using the one-loop matching condition \cite{HiggsPole}.

According to Eq. (\ref{nflag}) there are additional contributions to the one and two loop beta function for $\lambda$  which are proportional to the $\kappa_1$
and $\kappa_2$ couplings. At one loop we have
\begin{eqnarray}
 \delta\beta_{\lambda}^{(1)} = 12(\kappa_1^2+\kappa_2^2)\lambda -12(\kappa_1^4+\kappa_2^4),
\label{kappa+1}
\end{eqnarray}
and for two loop
\begin{eqnarray}
 \delta\beta_{\lambda}^{(2)}&=&\left(\frac{8}{5}g_1^2
-64 g^2_3\right)(\kappa_1^4+\kappa_2^4)-\frac{9}{2}g^4_2(\kappa_1^2+\kappa_2^2)+10\lambda\left(\frac{1}{4}g_1^2+\frac{9}{4}g_2^2+8g^2_3\right)(\kappa_1^2+\kappa_2^2) \nonumber \\
 && + \frac{3}{5} g_1^2\left(\frac{3}{2} g_1^2+9g^2_2\right)(\kappa_1^2+\kappa_2^2) -72\lambda^2(\kappa_1^2+\kappa_2^2)-3\lambda(\kappa_1^4+\kappa_2^4) + 60(\kappa_1^6+\kappa_2^6).
\label{kappa+2}
\end{eqnarray}

We next analyze the two loop RGEs numerically and show how
 the vacuum stability and perturbativity bounds on the SM Higgs boson mass
 are altered in the presence of the new TeV scale vectorlike particles.

 We chose the cutoff scale to be $M_{\rm GUT}$, the scale at which the SM gauge couplings are all equal. This choice is motivated by the following argument. Namely, we want to have as much as possible  model independent analysis
 and in the realistic GUT's  we can  have very  different representation for fields. For instance there are many choice of fileds  to break  GUT symmetry \cite{Nath:2006ut},  or  if one address the question of flavor structure of fermions in the framework of GUT, or origin of  neutrino  mass and  etc. Also  it is well known that in many GUT the  cutoff scale has to be  very close to the $M_{\rm GUT}$ scale doe to existence of big representation under the GUT gauge symmetry, for instance in SO(10), E(6) etc.

 We define the vacuum stability bound as the lowest Higgs boson mass
 obtained from the running of the Higgs quartic coupling
 which satisfies the condition $\lambda(\mu) \geq 0$,
 for any scale between $M_Z \leq \mu \leq M_{\rm GUT}$.
On the other hand, the perturbativity bound is defined as
 the highest Higgs boson mass obtained from the running
 of the Higgs quartic coupling with the condition
 $\lambda(\mu) \leq 4 \pi$ for any scale
 between $M_Z \leq \mu \leq M_{\rm GUT}$.

In Figure 1, we present the evolution of the gauge couplings for the SM (left panel) and for the extended SM (ESM) containing the vectorlike fermions $Q+\bar Q+D+\bar D$  (right panel).
As noted in  \cite{Amaldi:1991zx}, in ESM model with new vectorlike  fermions weighing a 100 GeV or so, one can realize essentially
perfect gauge coupling unification at some scale $M_{\rm GUT}$. Furthermore, if we require gauge coupling unification  at a level of around  $1\%$ or so, then the new vectorlike fermion mass should weigh less than a TeV. For definiteness, we set $M_{F}=500$ GeV in our calculation.
In this case the SM gauge couplings are unified at $M_{\rm GUT} \simeq 3 \times 10^{16}$ GeV.
As seen in Figure 1, the new vectorlike particles help achieve unification by altering the slopes of the three gauge couplings. In particular, the slope of  $\alpha_3$ is changed and it becomes larger at $M_{\rm GUT}$ in comparison to the SM case. The evolution of the top Yukawa coupling is also affected and its value is somewhat smaller at $M_{\rm GUT}$.

\begin{figure}[t]
\centering
\includegraphics[angle=0, viewport = -10 -10 240 200,width=8cm,]{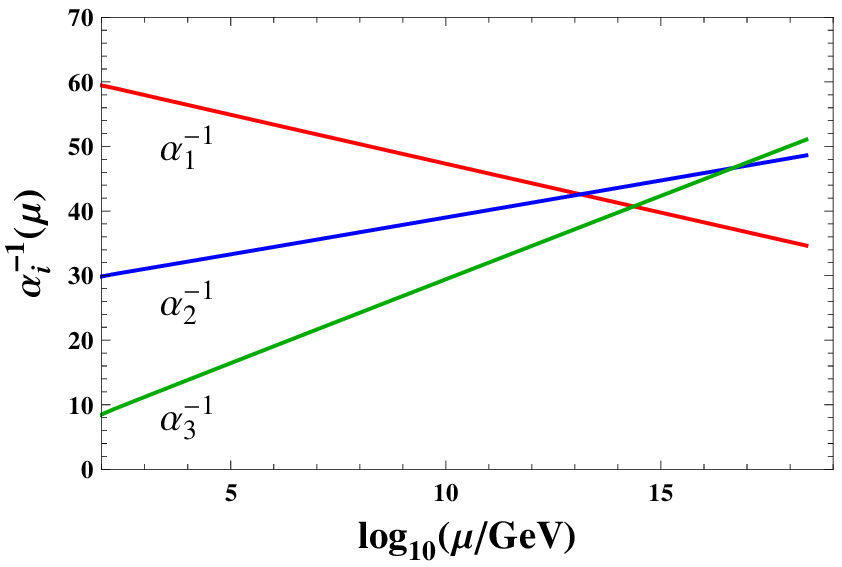}
\includegraphics[angle=0, viewport = -10 -10 240 200,width=8cm]{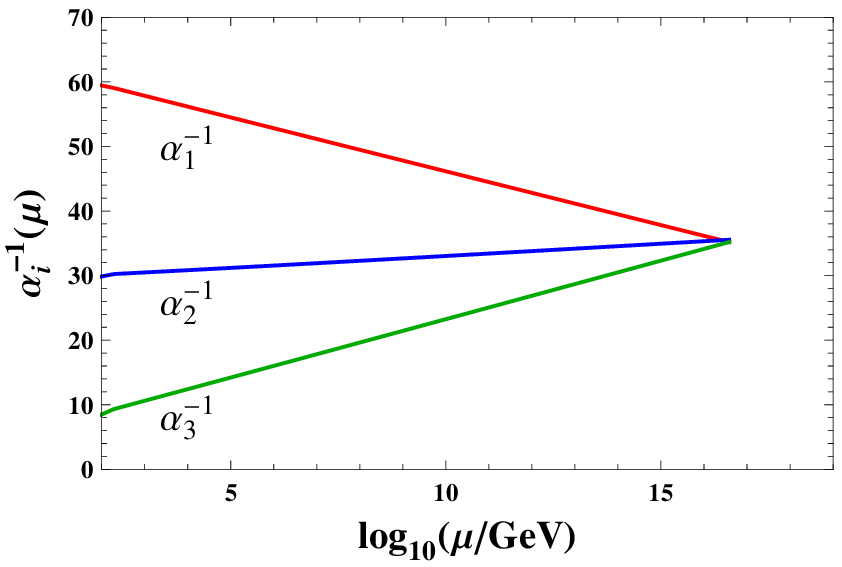}
\caption{\small
Gauge coupling evolution in the SM (left panel) and in  the extended SM (right panel).  The vectorlike mass is set equal to 500 GeV and the gauge coupling unification scale is $M_{GUT} \simeq 3 \times 10^{16}$ GeV.
 } \label{f1}
\end{figure}

\begin{figure}[t, width=12cm, height=10cm]
\begin{center}
{\includegraphics[height=7cm]{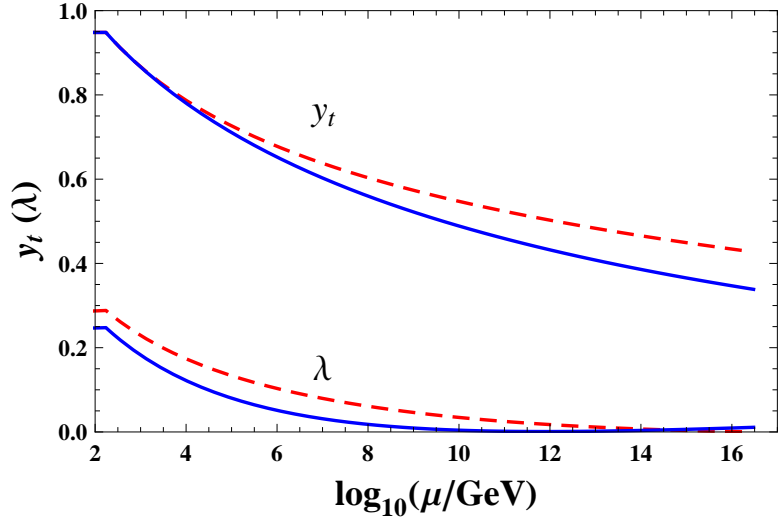}}
\end{center}
\caption{\small
  Evolution of the top Yukawa coupling in the SM (red dashed line) and in the extended SM (blue solid line). The evolution of the SM Higgs quartic coupling  in the two cases are also displayed. We have set $M_{F}=500$ GeV and $\kappa_i=0$. }
%
\end{figure}

 In Figure 2 we show how the evolution of the two-loop top Yukawa coupling in ESM with $M_{F}=500$ GeV. The red dashed line stands for the SM case, and the blue solid  line corresponds to the ESM with $\kappa_i=0$.
 We also present in Figure 2 the evolution of the Higgs quartic coupling. The red dashed line corresponds to the vacuum stability bound for Higgs quartic coupling in the SM, and the blue solid line corresponds to the quartic couplings in the ESM.
 We see that at $M_{\rm GUT}$, the top Yukawa coupling in the ESM  is smaller in comparison to the SM case. On the other hand, it is well known that  in  the determination of  the SM Higgs  boson mass vacuum stability  bound \cite{stability1}, a crucial role is played by the interplay between the top Yukawa coupling and Higgs quartic coupling, which have comparable and dominant contributions in the RGE for Higgs quartic coupling (See Eq. (\ref{self-1})). The negative sign contribution from the top Yukawa coupling  makes the Higgs quartic coupling smaller during the evolution. This is how the lower bound for Higgs boson mass is obtained in the SM. So, having in the model a smaller value  at $M_{\rm GUT}$ for the top Yukawa coupling means having a milder contribution in the RGE for the Higgs  quartic coupling, and this explains why in ESM, somewhat smaller values for the Higgs quartic coupling\footnote{ A similar observation was made in  ref. \cite{Gogoladze:2008ak} when considering the type III seesaw mechanism for neutrinos.} can satisfy the vacuum stability bound, compared to the SM.
In ESM, the lower bound for the SM Higgs boson mass using the one-loop matching condition \cite{HiggsPole} is found to be $m_H=117$ GeV, close to the LEP  bound of $114.4$ GeV~\cite{LEP2}. We estimate a theoretical error in this prediction of about 2 GeV, which is in addition to the errors arising from the experimental uncertainties in the determination of the top quark mass and $\alpha_3$ \cite{Espinosa:2007qp}.

\begin{figure}[t, width=12cm, height=10cm]
\begin{center}
{\includegraphics[height=7cm]{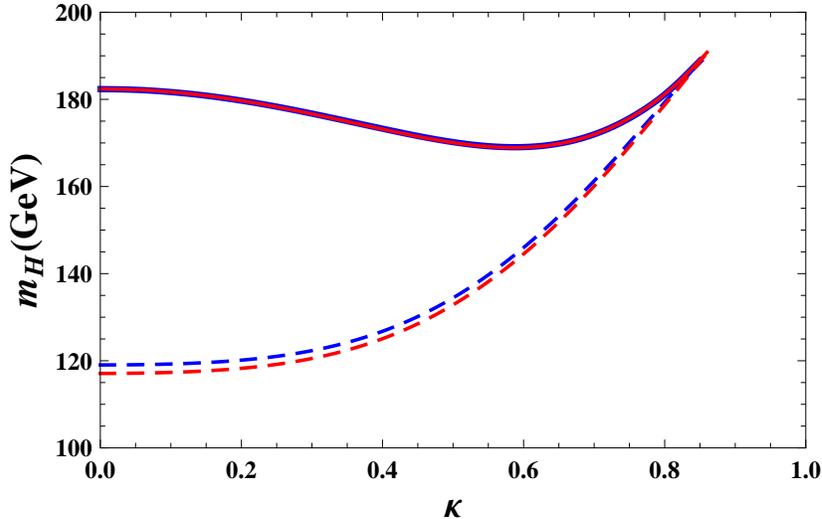}}
\end{center}
\caption{\small
  Perturbativity (solid) and vacuum stability (dashed) bounds on the Higgs boson pole mass ($m_H$) versus $\kappa(\equiv\kappa_1=\kappa_2)$, with vectorlike particle mass
  $M_{F}=500$ GeV (red lines) and  $M_{F}=1$ TeV (blue lines).
  The maximum value for the perturbativity bound  is $m_H\simeq191$ GeV  when $\kappa=0.86$.
  The lower bound for the Higgs mass is $m_H\simeq117$ GeV, with $\kappa=0$ and $M_{F}=500$ GeV. }
%
\end{figure}

As mentioned earlier, the $\kappa_i$ coupling in  Eq. (\ref{nflag}) can be  $O$(1) if $M_{F}> 500$ GeV.  In Figure 3 we
present the  Higgs boson mass versus  $\kappa_i$ for varying $M_{F}$ scales. For simplicity, we assume that  $\kappa \equiv \kappa_1=\kappa_2$. The upper solid blue and red curves
correspond to the Higgs perturbativity bound, and the lower dashed curves correspond to the vacuum stability bound when  the vectorlike particle mass is taken to be  500 GeV (dashed red) and 1 TeV (dashed blue).
 It is interesting to observe that the perturbativity  bound decreases as $\kappa$ increases from zero to $\kappa\approx 0.6$, and then increases as the value of $\kappa$  is increased further. We can easily understand this behavior at one loop level.   It arises from the interplay between the terms $12\lambda(\kappa_1^2+\kappa_2^2)$ and $-12(\kappa^4_1+\kappa^4_2)$ in Eq. (\ref {kappa+1}).
 Up to $\kappa\approx 0.6$, the term proportional to $\kappa^2\lambda$ dominates over the  $\sim \kappa^4$ contribution. So, for $\kappa \lesssim 0.6$, in the RGE in Eq. (\ref{self5}), we have an effective additional contribution with the same sign  as the $\lambda$ coupling, which leads to the decrease of the perturbativity bound. For $\kappa \gtrsim 0.6$ the $\sim \kappa^4$ contribution dominates compared to the term  $\kappa^2\lambda$, and we have an effective additional contribution which has  the same sign contribution  as the top quark in Eq. (\ref{self5}). This  leads to an increasing perturbativity bound as the $\kappa$ coupling increases. Note that we have an upper bound $\kappa=0.86$ for $M_{F}=500$ GeV, and  $\kappa=0.84$ for $M_{F}=1$ TeV. This happens because either the top Yukawa or $\kappa$ coupling becomes nonperturbative before the GUT scale. Corresponding to the  upper bound for $\kappa$ couplings, we have an upper bound on the Higgs mass: $m_H=191$ GeV if $M_{F}=500$ GeV, and $m_H=189$ GeV if $M_{F}=1$ TeV.

 We see in Figure 3 that the vacuum stability bound gradually increases as the $\kappa$ coupling increases. This happens because in the evolution of the Higgs quartic coupling, corresponding to the vacuum stability bound, the contribution proportional to the term $-\kappa^4$ dominates over the $\kappa^2\lambda$ contribution for lower values of $\kappa$. So in the RGE  for the Higgs quartic coupling (see Eq. (\ref{self5})) we have an additional  contribution with the same  sign  as the top quark. This leads to the explanation why the vacuum stability  bound increases when value of  $\kappa$ increases at low scale, and they eventually merge with the vacuum stability bound.
 We obtain the following results for the Higgs mass  corresponding to the vacuum stability bound: $m_H=117$ GeV when $M_{F}=500$ GeV, and  $m_H=119$ GeV when $M_{F}=1$ TeV, with $\kappa=0$.

\section{Type I Seesaw and the Higgs Boson Mass}

We next consider the impact of type I seesaw physics \cite{seesawI} on the Higgs mass bounds found in the previous section.
The terms relevant for neutrino oscillations through type I seesaw are given by
\begin{eqnarray}
{\cal L}_{\nu}= -y_D^{ij}\, l_i\, \nu^c_j\, \Phi^c- \frac{1}{2}M_R^{ij}(\nu^c)^T_i\nu_j + h.c.,~~~i,j=1,2,3.
\label{lagsees}
\end{eqnarray}
\begin{figure}[t, width=12cm, height=8cm]
\begin{center}
{\includegraphics[height=7cm]{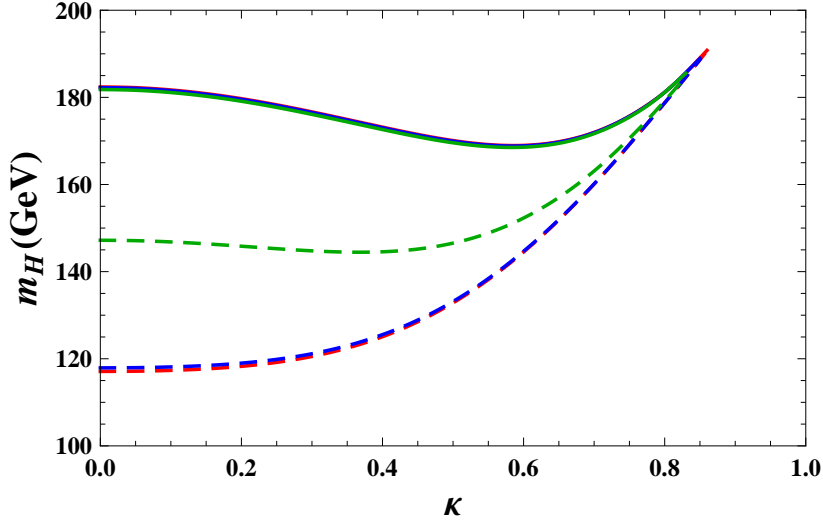}}
\end{center}
\caption{\small
  Perturbativity (solid) and vacuum stability (dashed) bounds on the Higgs boson pole mass ($m_H$) versus $\kappa(\equiv\kappa_1=\kappa_2)$ in the extended SM, including type I seesaw physics. We consider three different type I seesaw scales $M_R=10^{13}$ GeV (red), $10^{14}$ GeV (blue) and $10^{15}$ GeV (green).  For our calculation we consider a hierarchical neutrino mass spectrum, and we set $M_{F}=500$ GeV.
  The maximum and minimal values  for the Higgs mass corresponding to the perturbativity and vacuum stability bounds
  are the same as in Figure 3.}
%
\end{figure}
Here $l_i$ is the lepton doublet,  $\nu^c_{i}$ the right handed neutrino,  $y_D^{ij}$ is  neutrino Yukawa coupling and $M_{R}^{ij}$ denotes the right handed neutrino mass matrix.

Above the scale $M_R$ we have the following one loop RGE for $Y_{\nu}\equiv y_D^{ij}$,
\begin{eqnarray}
\frac{d {\bf Y_\nu}}{d \ln \mu}
 =\frac{1}{16 \pi^2} {\bf Y_\nu}
  \left( 3 y_t^2 +  {\rm tr}\left[ Y_\nu^{\dagger}Y_\nu \right]+  \frac{3}{2} Y_\nu^{\dagger}Y_\nu
   -\left( \frac{9}{20} g_1^2 +\frac{9}{4} g_2^2 \right) \right).
\label{lagsees1}
\end{eqnarray}
The various beta functions are modified as follows:
\begin{eqnarray}
&& \beta_t^{(1)} \to  \beta_t^{(1)}
 + {\rm tr}\left[ Y_\nu^{\dagger}Y_\nu  \right], \nonumber \\
 && \beta_{\kappa_1}^{(1)} \to  \beta_{\kappa_1}^{(1)}
 + {\rm tr}\left[ Y_\nu^{\dagger}Y_\nu  \right], \nonumber \\
 && \beta_{\kappa_2}^{(1)} \to  \beta_{\kappa_2}^{(1)}
 + {\rm tr}\left[ Y_\nu^{\dagger}Y_\nu  \right], \nonumber \\
&&  \beta_\lambda^{(1)} \to
  \beta_\lambda^{(1)}
 + 4 \; {\rm tr}[Y_\nu^{\dagger}Y_\nu ] \lambda
 - 4 \; {\rm tr}[(Y_\nu^{\dagger}Y_\nu )^2] .
\label{RGE-type5}
\end{eqnarray}

It is certainly interesting to consider  realistic cases of the neutrino mass matrix and mixing
 which reproduce the current neutrino oscillation data. We will consider a scenario in which the light neutrinos form a hierarchical mass spectrum.
 It was shown in Ref. \cite{Gogoladze:2008ak} that the impact on the SM Higgs boson mass from an inverted-hierarchial neutrino mass spectrum is not significantly different from the hierarchial case.

The light neutrino mass matrix is diagonalized
 by a mixing matrix $U_{MNS}$ such that
\begin{eqnarray}
  {\bf M}_\nu = \frac{v^2}{2 M} \; Y^T Y
   = U_{MNS} D_\nu U^T_{MNS},
\end{eqnarray}
with $D_\nu ={\rm diag}(m_1, m_2, m_3)$,
 where we have assumed, for simplicity, that
 the Yukawa matrix ${\bf Y}_\nu$ is real.
We further assume that the mixing matrix has
 the so-called tri-bimaximal form \cite{hps}
\begin{eqnarray}
U_{MNS}=
\left(
\begin{array}{ccc}
\sqrt{\frac{2}{3}} & \sqrt{\frac{1}{3}} & 0 \\
-\sqrt{\frac{1}{6}} & \sqrt{\frac{1}{3}} &  \sqrt{\frac{1}{2}} \\
-\sqrt{\frac{1}{6}} & \sqrt{\frac{1}{3}} & -\sqrt{\frac{1}{2}}
\end{array}
\right) ,
\label{ansatz}
\end{eqnarray}
 which is in very good agreement with the current
 best fit values of the neutrino oscillation data \cite{NuData}.

For the hierarchical case the diagonal neutrino mass matrix is given by
\begin{eqnarray}
 D_\nu \simeq
 {\rm diag}(0,\sqrt{\Delta m_{12}^2}, \sqrt{\Delta m_{23}^2}).
\end{eqnarray}
 We fix the input values for the solar and atmospheric
 neutrino oscillation data as \cite{NuData}
\begin{eqnarray}
 \Delta m_{12}^2 &=& 8.2 \times 10^{-5} \; {\rm eV}^2,\nonumber \\
 \Delta m_{23}^2 &=& 2.4 \times 10^{-3} \; {\rm eV}^2.
 \label{massdiff}
\end{eqnarray}

Our finding are  presented in Figure 4 where we plot the vacuum stability (dashed) and perturbativity (solid) bound  versus $\kappa(\equiv\kappa_1=\kappa_2)$, with $M_{F}$ set equal to 500 GeV.  We consider three distinct mass scales for the heavy right handed neutrinos, namely, $M_R=10^{13}$ GeV (red), $10^{14}$ GeV (blue) and $10^{15}$ GeV (green). The general picture of the Higgs mass versus $\kappa$ coupling is qualitatively the same as in Figure 3. Only the initial values for the Higgs mass  when $\kappa=0$ is taken are changed depending on the type I seesaw scale.
 According to Eq. (\ref{RGE-type5}) the Dirac neutrino Yukawa coupling $Y_{\nu}$ gives an additional contribution to the Higgs quartic coupling RGE with the same sign as the top quark contribution. It is natural to expect that the vacuum stability bound will increase if the $Y_{\nu}$ coupling is increased. For  $M_R=10^{13}$ GeV, the vacuum stability bound essentially coincides with the corresponding bounds in Figure 3. With $M_R=10^{14}$ GeV the vacuum stability bound is only slightly altered since $Y_{\nu}$ is still not large at that scale in comparison to the top Yukawa coupling. For  $M_R=10^{15}$ GeV we see a significant change in the
 vacuum stability bound since now the coupling $Y_{\nu}$ is larger than the top Yukawa coupling, and the two of them together force the Higgs quartic coupling at low scale to be larger in order to satisfy the vacuum stability bound. Note that there is hardly any impact of type I seesaw on the perturbativity bound. This is due to the fact that above the seesaw scale the Higgs quartic coupling is already larger than  $Y_{\nu}$.

\section{Conclusion}

Following ref. \cite{Amaldi:1991zx}, we have considered a plausible extension of the SM in which new vectorlike fermions  carrying SM quantum numbers and with masses of order 300 GeV -- 1 TeV are introduced. This relatively modest extension of the SM, denoted by ESM in the text, leads to a rather precise unification of the SM gauge couplings at $M_{\rm GUT}\sim 3\times 10^{16}$ GeV, and it also gives rise to a vacuum stability bound on the SM Higgs mass of 117 GeV. The perturbativity bound on the Higgs mass is estimated to lie close to 190 GeV. The new vectorlike fermions should be accessible at the LHC.

\section*{Acknowledgments}

We thank Yukihiro Mimura,  Nobuchika Okada, Mansoor Ur Rehman  and Joshua R. Wickman for useful comments and discussion.  This work is supported in part by the DOE Grant
\#DE-FG02-91ER40626 (I.G. H.B. and Q.S.), GNSF grant 07\_462\_4-270 (I.G.) and by Bartol Research Institute (B.H.).

\end{document}